\documentclass[pra,reprint,twocolumn,superscriptaddress,showpacs,floatfix]{revtex4-1}
\usepackage{amsmath}
\usepackage{mathrsfs}
\usepackage{txfonts}
\usepackage{amssymb}
\usepackage{graphicx}
\usepackage{hyperref}
\usepackage{ulem}
 \usepackage{overpic}
 \usepackage{psfrag}
\usepackage{tabularx}
\usepackage{array}
\usepackage{placeins}
\usepackage{xcolor}
\newcommand{\PreserveBackslash}[1]{\let\temp=\\#1\let\\=\temp}

\makeatletter

\begin{document}

\title{Entanglement entropy for a type of scale-invariant states in two spatial dimensions and beyond: universal finite-size scaling}

\author{Huan-Qiang Zhou}
\affiliation{Centre for Modern Physics, Chongqing University, Chongqing 400044, The People's Republic of China}

\author{Qian-Qian Shi}
\affiliation{Centre for Modern Physics, Chongqing University, Chongqing 400044, The People's Republic of China}

\author{Ian P. McCulloch}
\affiliation{Department of Physics, National Tsing Hua University, Hsinchu 30013, Taiwan}
\affiliation{Frontier Center for Theory and Computation, National Tsing Hua University, Hsinchu 30013, Taiwan}

\author{Murray T. Batchelor}
\affiliation{Mathematical Sciences Institute, The Australian National University, Canberra ACT 2601, Australia}
\affiliation{Centre for Modern Physics, Chongqing University, Chongqing 400044, The People's Republic of China}

\begin{abstract}
A generic scheme is proposed to investigate the entanglement entropy for a type of scale-invariant states, valid for orthonormal basis states in the ground state subspace of quantum many-body systems undergoing spontaneous symmetry breaking with type-B Goldstone modes in two spatial dimensions and beyond. It is argued that a contribution from the area law to the entanglement entropy is absent, since the closeness to the boundary between a subsystem and its environment is not well-defined, given that a permutation symmetry group with respect to the unit cells of degenerate ground state wave functions emerges. Three physical constraints imposed lead to a universal finite-system size scaling function in the dominant logarithmic contribution to the entanglement entropy. As a result, an abstract fractal underlying the ground state subspace is revealed, characterized by the fractal dimension. The latter in turn is identical to the number of type-B Goldstone modes for the  orthonormal basis states. The prediction is numerically confirmed for the ${\rm SU}(2)$ spin-$s$ ferromagnetic Heisenberg model, the ${\rm SU}(2s+1)$ ferromagnetic model, and the staggered ${\rm SU}(3)$ spin-1 ferromagnetic biquadratic model.
\end{abstract}.

\maketitle

\onecolumngrid

\section{Introduction}

Over the last decades, effort has been made in an attempt to classify distinct types of the Goldstone modes (GMs)~\cite{goldstone,Hnielsen, schafer, miransky, nambu, nicolis,  brauner-watanabe, watanabe, NG}, which emerge as a result of spontaneous symmetry breaking (SSB) for continuous symmetry groups.  Historically, this may be traced back to the pioneering work by Goldstone, who discovered the emergence of a gapless mode if a continuous symmetry group $G$ is broken into a residual symmetry group $H$~\cite{goldstone}. Recent developments reveal that a proper distinction must be made between type-A GMs and type-B GMs, thus leading to the establishment of the counting rule for  GMs~\cite{watanabe,NG}: $N_A+2N_B=N_{BG}$,
where $N_A$ and $N_B$ are the numbers of  type-A and type-B GMs, and $N_{BG}$ is the number of broken generators, equal to the dimension of the coset space $G/H$.

In sharp contrast to SSB with type-A GMs, quantum many-body systems exhibiting SSB with type-B GMs are quite rare in condensed matter, although the familiar ${\rm SU}(2)$ ferromagnetic Heisenberg model offers a paradigmatic example. Indeed,
not until very recently have a few other examples for SSB with type-B GMs emerged in quantum many-body  systems in one spatial dimension~\cite{FMGM,LLspin1,golden,SU4}. An important question concerns a complete understanding of SSB with both type-A and type-B GMs from the perspective of quantum entanglement. Indeed, a systematic investigation into the entanglement entropy has been carried out for quantum many-body systems undergoing SSB with type-A GMs~\cite{Kallin,Song, Metlitski,squarecubic,bilayer,typeAGM} in two spatial dimensions and for those undergoing SSB with type-B GMs in one spatial dimension~\cite{FMGM,LLspin1,golden,SU4}. Originally,  a scaling relation of the entanglement entropy for SSB with type-A GMs in two and higher spatial dimensions, proposed in
Ref.~\cite{Metlitski}, was inspired by quantum Monte Carlo simulations~\cite{Kallin} and spin-wave calculations~\cite{Song}
for the two-dimensional ${\rm SU}(2)$ spin-$1/2$ Heisenberg antiferromagnetic model.
However, the current understanding is still limited, since a universal finite system-size scaling function is lacking.

As we have learned from the known examples, quantum many-body systems exhibiting SSB with type-B GMs are described by the so-called frustration-free Hamiltonians~\cite{tasaki,katsura}. They are always exactly solvable, as far as highly degenerate ground states are concerned. Indeed, some of them are even completely integrable in the Yang-Baxter sense, if only considering their one-dimensional versions, given that the models may be associated with a solution to the quantum Yang-Baxter equation~\cite{baxterbook,Sutherland}. In fact, there are examples which appear as a representation of the Temperley-Lieb algebra~\cite{tla,baxterbook,martin}. A notable feature is that highly degenerate ground states arising from SSB with type-B GMs are
scale-invariant, but not conformally invariant, in one spatial dimension. Actually, they exhibit different universal finite-size scaling behaviours in the entanglement entropy, thus making it possible to distinguish between scale-invariant states and conformally invariant states~\cite{finitesize}.

It is thus highly desirable to propose a generic scheme to perform a finite system-size scaling analysis of the entanglement entropy for quantum many-body systems undergoing SSB with type-B GMs in two spatial dimensions and beyond. For this purpose, two questions need to be addressed. The first question concerns a possible contribution from the area law ~\cite{srednicki,Eisert} to the entanglement entropy. Here, the entanglement entropy is defined if the system is partitioned into a block and its environment. As it turns out, this contribution is absent for almost all known quantum many-body systems exhibiting SSB with type-B GMs, because the closeness of degrees of freedom to the boundary between a subsystem and its environment and the boundary itself are  not well-defined, given the emergence of a permutation symmetry group with respect to the unit cells of degenerate ground state wave functions arising from SSB with type-B GMs. Indeed, the permutation symmetry group for these types of scale-invariant states is {\it emergent}, since the model Hamiltonian itself is not permutation-invariant. The second question concerns the remaining dominant contribution to the entanglement entropy, expressed as a logarithm of a universal finite system-size scaling function, with the constraint imposed on this scaling function that it reduces to the block size in the thermodynamic limit. 

In order to determine such a universal  finite system-size scaling function,  it is necessary to further develop our previous argument~\cite{finitesize}, devised for quantum many-body systems undergoing SSB with type-B GMs in one spatial dimension. The key ingredients may be formulated as the following three physical constraints: first, it must be identical for a block and for its environment for highly degenerate ground states as pure quantum states;  second, it must tend to be the block size if the thermodynamic limit is approached; third, as a homogeneous function of both the system size and the block size, it is linear in each argument, to render consistency with a scale transformation.
As it turns out, the  universal finite-size scaling function for a type of scale-invariant states arising from SSB with type-B GMs is the
most fundamental solution satisfying the above three constraints. We remark that this also sheds light on our search for a  universal finite-size scaling function in a sub-leading logarithmic correction to the leading contribution from the area law for scale-invariant states arising from SSB with type-A GMs.

The layout of this paper is as follows. In Section~\ref{absence}, we argue that the contribution from the area law to the entanglement entropy is absent for scale-invariant states arising from SSB with type-B GMs, if they are permutation-invariant. In Section~\ref{universalscaling}, we present a generic scheme to  establish a universal finite system-size scaling function. In Section~\ref{threeexamples}, we explicitly construct the orthonormal basis states
for three fundamental models exhibiting SSB with type-B GMs. These are the ${\rm SU}(2)$ spin-$s$ ferromagnetic Heisenberg model, the ${\rm SU}(2s+1)$ ferromagnetic model, and the staggered ${\rm SU}(3)$ spin-1 ferromagnetic biquadratic model. In Section~\ref{ee}, a numerical test is performed  to confirm the theoretical prediction for the three illustrative models. The last Section~\ref{summary} is devoted to a summary.

\section{Absence of the contribution from the area law to the entanglement entropy for scale-invariant states arising from SSB with type-B GMs: permutation invariance} \label{absence}

Consider a  quantum many-body system, described by the Hamiltonian $\mathscr{H}$, on a two-dimensional lattice. Suppose it undergoes SSB from $G$ to $H$, with type-B GMs. Here $G$ denotes a (semisimple) symmetry group and $H$ denotes a residual symmetry group. For simplicity, we restrict ourselves to a square lattice, with the two linear sizes being $L_1$ and $ L_2$.
If the system is partitioned into a block $B$ and its environment $E$, with the block consisting of $n_1\times n_2$ (contiguous) lattice sites, and the environment $E$ consisting of the other $L_1L_2-n_1n_2$ lattice sites, then one may introduce the reduced density matrix $\rho(L_1,L_2,n_1,n_2)$ for the block, which in turn allows us to define
the entanglement entropy $S(L_1,L_2,n_1,n_2)$. Mathematically, it is defined as $S(L_1,L_2,n_1,n_2)=-{\rm Tr}(\rho(L_1,L_2,n_1,n_2)\log_2 \rho(L_1,L_2,n_1,n_2))$.

If $L_1$, $L_2$, $n_1$ and $n_2$  are finite but large enough,  one may anticipate that the entanglement entropy $S(L_1L_2,n_1n_2)$ consists of the contributions from  three distinct parts, if one only considers a smooth boundary  between the block and the environment. The first contribution $a {\cal A}$ comes from the requirement by the area law~\cite{srednicki,Eisert}, where $\cal A$ denotes the area of the boundary between the block and the environment, with the coefficient $a$ being non-universal. The second contribution is logarithmic and
takes the form $\log_2 g(L_1,L_2,n_1,n_2)$, where $g(L_1,L_2,n_1,n_2)$ is introduced as a universal finite system-size scaling function, yet to be determined. Meanwhile, the physical meaning of a universal finite system-size scaling function needs to be clarified. The third contribution is an additive non-universal constant $c$, originating from model-dependent factors. Here, it is proper to stress that the discussion up to now is applicable to quantum many-body systems undergoing SSB with type-A or type-B GMs.
Indeed, as demonstrated in Ref.~\cite{Metlitski} (cf. also Refs.~\cite{typeAGM}) for SSB with type-A GMs,  the first (leading) contribution plays an essential role in quantifying quantum entanglement present in the unique ground state as a singlet of the symmetry group $G$, in order to keep consistency with the area law~\cite{srednicki,Eisert}. Instead, if there is a corner in the  boundary  between the block and the environment, then a corner contribution appears for  SSB with type-A GMs, as already discussed in Ref.~\cite{Metlitski}. In contrast, as we shall argue below, no corner contribution exists for SSB with type-B GMs.

A notable feature for quantum many-body systems in two spatial dimensions, which undergo SSB with type-B GMs, is that the leading non-universal contribution from the area law is absent, if degenerate ground states are permutation-invariant with respect to the {\it emergent} unit cells, which may be or may not be identical to the lattice unit cells. Here, we stress that permutation invariance is valid for almost all known quantum many-body systems exhibiting SSB with type-B GMs~\cite{FMGM,LLspin1,golden,SU4}, with the  flat-band ferromagnetic Tasaki model as a notable exception, although it exhibits SSB  from ${\rm SU}(2)$ to ${\rm U}(1)$ with one type-B GM~\cite{TypeBtasaki}.
In fact, the emergent unit cells are explicitly present in exact matrix product state (MPS) representations for the orthonormal basis states. Actually, it is straightforward to extend the construction of an exact MPS representation for  the orthonormal basis states, developed in Ref.~~\cite{exactmps}, to quantum many-body systems undergoing SSB with type-B GMs in two spatial dimensions (cf. Sec. A of the Supplementary Material (SM) for more details).
In fact, as argued in Ref.~\cite{srednicki}, the entanglement entropy should depend only on properties shared by a given block and its environment, i.e., the boundary between them. It follows that only the degrees of freedom located in the vicinity of the boundary contribute. However, the emergence of the permutation invariance with respect to the  emergent unit cells implies that the number of degrees of freedom located in the vicinity of the boundary between the block and the environment is not well-defined, as one sees from the outcomes of permutation operations restricted to the block and the environment and that the boundary itself is not well-defined if a permutation symmetry operation involving the lattice sites in both the block and the environment is performed. In other words, the closeness to a boundary between a block and its environment and the boundary itself are not well-defined for a permutation-invariant ground state, depending on the type of a permutation operation involved. This leads to an apparent contradiction to the area law, if the leading contribution exists. Hence the only way out is to require that the coefficient $a$ be zero. That is, the sub-leading universal logarithmic contribution to the entanglement entropy dominates for quantum many-body systems undergoing SSB with type-B GMs, as long as degenerate ground states are permutation-invariant. In other words, the dominant contribution to the entanglement entropy thus comes from a universal logarithmic scaling term for permutation-invariant (degenerate) ground states. Meanwhile,  the permutation invariance implies that no corner contribution exists for SSB with type-B GMs, since the boundary itself is not well-defined.

We stress that the same argument is valid for quantum many-body systems undergoing SSB with type-B GMs in higher spatial dimensions. The absence of the leading non-universal contribution from the area law to the entanglement entropy implies that highly degenerate ground states are far less entangled than one might have expected in two and higher spatial dimensions, though usually they are much more entangled than critical ground states described by conformal field theory in one spatial dimension~\cite{cft}. Actually, the presence of exact MPS representations for  the orthonormal basis states in the ground state subspace of quantum many-body systems undergoing SSB with type-B GMs in two and higher spatial dimensions (cf. Sec. A of the SM) lends further support to the claim that they escape from the constraint of the area law regardless of the spatial dimensionalities, as far as the entanglement entropy for a permutation-invariant ground state is concerned. As already mentioned above, the permutation invariance for degenerate ground states is present for quantum many-body spin systems investigated in Refs.~\cite{FMGM,LLspin1,golden, finitesize, SU4, dtmodel,cantorset}, but the  flat-band ferromagnetic Tasaki model is an exception~\cite{TypeBtasaki}. Hence  it is necessary to further investigate the entanglement entropy for its two-dimensional version, which is a paradigmatic example for SSB with type-B GMs as an itinerant electron model.  

This marks a drastic difference between type-A and type-B GMs, in addition to a well-known difference that a quantum many-body system undergoing SSB with type-A GMs entails the Anderson tower of states, in contrast to degenerate ground states in a quantum many-body system undergoing SSB with type-B GMs.  For the former, the entanglement entropy is usually evaluated for the unique ground state, as a singlet of the symmetry group $G$, whereas for the latter,  the entanglement entropy is defined for each of highly degenerate ground states, if the system size is finite.  Actually, there exist two energy scales associated with a type-A GM, i.e., the gap in  the Anderson tower of states: $\Delta \sim L_g^{-d}$ and the gap relevant to a type-A GM: $\Delta_{gm}\sim L_g^{-1}$, with $d$ being the spatial dimension. In contrast, there emerge two energy scales associated with a type-B GM, which are related with the (spatial) correlation length and (temporal) coherence time:  $\Delta_{gm}\sim L_g^{-z}$ and $\Delta_{gm}\sim L_g^{-1}$ , with $z$ being the dynamical critical exponent. Here $L^d_g$ is defined as $L^d_g=L_1L_2 \ldots L_d$, with $L_\delta$ ($\delta =1, \ldots,d$) being the linear sizes.

The peculiarity of SSB with type-B GMs lies in the fact that the number of degrees of freedom, which contribute to the entanglement entropy, may be counted as the spatial fine structure of the entanglement entropy. Mathematically, it is quantified in terms of the so-called entanglement contour~\cite{guifre}, and is subject to a simple characterization. Indeed, the entanglement contour for  a quantum many-body system undergoing SSB with type-B GMs is flat, in the sense that each of the unit cells in a chosen block $B$ contributes equally to the entanglement entropy as a result of the permutation symmetry subgroup restricted to the block itself. In other words, quantum many-body systems undergoing SSB with type-B GMs constitute  a realization of the simplest spatial fine structure from a  perspective of the entanglement contour (cf.~Sec.~B of the SM for more details). As a consequence,  the entanglement entropy $S(L_1,L_2,n_1,n_2)$ and the universal finite-size scaling function $g(L_1,L_2,n_1,n_2)$ depend on $L_1$, $L_2$, $n_1$ and $n_2$  through $L_1L_2$ and $n_1n_2$ in two spatial dimensions. We thus have $S(L_1,L_2,n_1,n_2)\equiv S(L_1L_2,n_1n_2)$ and  $g(L_1,L_2,n_1,n_2) \equiv g(L_1L_2,n_1n_2)$. Needless to say, this conclusion is also valid for the entanglement entropy. An alternative way to justify this dependence comes from an observation that it is the only choice to keep consistency with the possibility for deforming a lattice in $d$ spatial dimensions to a chain in one spatial dimension (cf.~Sec.~A of the SM).

\section{Universal finite system-size scaling functions in two spatial dimensions and beyond}~\label{universalscaling}

Here we present a generic scheme  to establish a universal finite system-size scaling function $g(L_1L_2,n_1n_2)$ for the orthonormal basis states, which appear as highly degenerate ground states arising from SSB with type-B GMs in a quantum many-body system in two and higher spatial dimensions.
Mathematically, the orthonormal basis states are constructed from the repeated action of the lowering operators $F_\alpha$ on the highest weight state or generalized highest weight states~\cite{FMGM,golden}.
In particular, for such a scale-invariant state, the dominant contribution to the entanglement entropy $S(n_1n_2)$ must exhibit a logarithmic scaling relation with the block size $n_1n_2$ in the thermodynamic limit:  $\log_2 g(L_1L_2,n_1n_2) \propto \log_2 (n_1n_2)$~\cite{FMGM}. Note that $S(n_1n_2)$ is defined as the thermodynamic limit of $S(L_1L_2,n_1n_2)$. Here by the thermodynamic limit we mean $L_1\rightarrow \infty$ and $L_2\rightarrow \infty$.

Three constraints emerge from the following three physical requirements, thus imposing  the restrictive conditions on $g(L_1L_2,n_1n_2)$.
First, for any pure quantum state, the entanglement entropy for a block $B$ is identical to that for its environment~\cite{nielsen}. Hence we have
\begin{equation}
g(L_1L_2,n_1n_2)=g(L_1L_2,L_1L_2-n_1n_2).
\label{con1}
\end{equation}
Second, as the thermodynamic limit is approached, one may expect that $g(L_1L_2,n_1n_2)$ becomes $n_1n_2$, i.e.
\begin{equation}
\lim\limits_{L_1,L_2\rightarrow \infty}g(L_1L_2,n_1n_2)\propto n_1n_2.
\label{con2}
\end{equation}
Hence the thermodynamic limit may be achieved if $L_1 \rightarrow \infty$ for fixed $L_2/L_1$ and vice versa.
Third, $g(L_1L_2,n_1n_2)$ must be a homogeneous function of $L_1L_2$ and $n_1n_2$, as a result of a scale transformation:  $L_1\rightarrow \lambda_1 L_1$, $L_2\rightarrow \lambda_2 L_2$, $n_1\rightarrow \lambda_1 n_1$,
$n_2\rightarrow \lambda_2 n_2$, given that it becomes $n_1n_2$ in the thermodynamic limit [cf.~Eq.~(\ref{con2})]. In other words,
$g(L_1L_2,n_1n_2)$ is bilinear in $L_1$, $n_1$ and $L_2$, $n_2$:
\begin{equation}
g(\lambda_1\lambda_2L_1L_2,\lambda_1\lambda_2n_1n_2)=\lambda_1\lambda_2g(L_1L_2,n_1n_2).
\label{con3}
\end{equation}

In principle, all possible universal finite system-size scaling functions must follow from solving the three constraints (\ref{con1}), (\ref{con2}) and (\ref{con3}), subject to other physical requirements. In particular, for fixed $L_1L_2$, $g(L_1L_2,n_1n_2)$ must be monotonically increasing with increasing $n_1n_2$ until $n_1n_2$ reaches $L_1L_2/2$.  Here we restrict ourselves to looking for the simplest solution to the three constraints, which turns out to be sufficient for describing  universal finite system-size scaling behaviors of the entanglement entropy
$S(L_1L_2,n_1n_2)$ for a quantum many-body system in two spatial dimensions, which undergoes SSB with type-B GMs.

Setting $\lambda_1=1/L_1$, $\lambda_2=1/L_2$ in the constraint (\ref{con3}), we have
\begin{equation}
g(L_1L_2,n_1n_2)=L_1L_2k \left(\frac{n_1n_2}{L_1L_2} \right).
\end{equation}
Here, $k(n_1n_2/L_1L_2)$ is a function of $n_1n_2/L_1L_2$, yet to be determined. That is, the problem is now reduced to the determination of the function $k(x)$, with $x=n_1n_2/L_1L_2$. Note that $k(x)$ must be monotonically increasing with increasing $x$ from 0 to $1/2$, and symmetric with respect to $x=1/2$.
	
The simplest form of $k(x)$ is $k(x)=u(x)u(1-x)$, with $u(x)=x$, in order to ensure that the first constraint (\ref{con1}) and the second constraint (\ref{con2}) are satisfied. Meanwhile, one has to retain consistency with  our previous results for the universal finite system-size scaling function of quantum many-body systems, undergoing SSB with type-B GMs, in one spatial dimension~\cite{FMGM,LLspin1,golden,SU4}.  In other words, we require that the entanglement entropy for quantum many-body systems in two spatial dimensions becomes that for quantum many-body systems in one spatial dimension, if one of the two linear sizes $L_1$ and $L_2$ is unity.
Here, we emphasize that this argument marks two remarkable differences between type-A and type-B GMs. The first is that {\it only} type-B GMs survive in one spatial dimension, and the second is that the emergent permutation symmetry makes it possible to turn degenerate ground states in two spatial dimensions into those in one spatial dimension.
Therefore, for a scale-invariant state in two spatial dimensions, the entanglement entropy $S_{\!\!f}(L_1L_2,n_1n_2)$ takes the form
\begin{equation}
	S_{\!\!f} (L_1L_2,n_1n_2)=\frac{N_B}{2} \log_2\frac{n_1n_2(L_1L_2-n_1n_2)}{L_1L_2} +S_{\!\!f0},
	\label{slnf}
\end{equation}
where $f$ is introduced to denote a set of fillings $f_1, f_2, \ldots, f_r$, with $r$ being the rank of the (semisimple) symmetry group $G$ for a quantum many-body system, and $S_{\!\!f0}$ denotes an additive non-universal constant. Note that $f_1, f_2, \ldots, f_r$ are defined relative to the highest weight state or a chosen generalized highest weight state. Mathematically, we have $f_1= M_1/(L_1L_2), f_2= M_2/(L_1L_2), \ldots, f_r= M_r/(L_1L_2)$, with $M_1, M_2,\ldots, M_r$ being the numbers of times to act the $r$ (commuting) lowering operators of $G$ on the highest weight state or a generalized highest weight state (cf. Sec. C of the SM for more details).

We stress that the scaling relation (\ref{slnf}) is universal, in the sense that the prefactor, denoted as $\kappa$, in front of the logarithmic function is linearly proportional to the number of type-B GMs $N_B$, as far as the orthonormal basis states are concerned, as already argued for one-dimensional quantum many-body systems undergoing SSB with type-B GMs~\cite{FMGM}. We remark that $\kappa$ must be a function of  the number of type-B GMs $N_B$, since only the low-lying excitations contribute to the scaling behaviour of the entanglement entropy $S_{\!\!f} (L_1L_2,n_1n_2)$. Hence, we have $\kappa = \kappa(N_B)$. Now imagine a fictitious system consisting of two arbitrary quantum many-body subsystems that are not coupled to each other, with the numbers of type-B GMs being $ N_{B,1}$ and $N_{B,2}$, respectively. Then, the total number of type-B GMs for the fictitious system is $N_B= N_{B,1}+N_{B,2}$. Accordingly, we have $\kappa(N_B)=\kappa( N_{B,1})+\kappa(N_{B,2})$, given that the entanglement entropy is additive for two subsystems that are uncoupled. This implies that $\kappa(N_B)$ is linearly proportional to $N_B$.
The proportionality constant may be determined from a specific model on a square lattice.  For the $\rm{SU(2)}$ spin-$1/2$ ferromagnetic states on a square lattice,  an analytic treatment yields that the  proportionality constant is $1/2$ for the  orthonormal basis states.

A few remarks are in order.
First,  we have assumed that the block consist of contiguous sites. However, this is not necessary, as a result of the emergent permutation symmetry for this type of scale-invariant states.
Second, the physical meaning of the prefactor in front of a logarithmic scaling function is that it is half the number of type-B GMs, which counts the number of gapless excitations in a scale-invariant state. 
This simply follows from an observation that a lattice in $d$ spatial dimensions may be deformed to a chain, as far as degenerate ground states are concerned (cf.~Sec.~A of the SM for an illustration in a two-dimensional square lattice).
Third, the scaling relation (\ref{slnf})  remains intact as the boundary conditions vary from open boundary conditions (OBCs) to  periodic boundary conditions (PBCs) for highly degenerate ground states arising from SSB with type-B GMs, but not for those arising from SSB with type-A GMs, as long as a degenerate ground state under OBCs remains to be a ground state under PBCs (though it is generically not necessarily the case). This originates from the flat entanglement contour for  SSB with type-B GMs. Another implication is that the origin of the logarithmic scaling term in the entanglement entropy for a quantum many-body system undergoing SSB with type-B GMs is different from the counterpart for  those undergoing SSB with type-A GMs in two and higher spatial dimensions, and for critical quantum many-body systems described by conformal field theory in one spatial dimension~\cite{cft}.
Fourth, it is straightforward to extend our argument to the entanglement entropy for scale-invariant states in higher spatial dimensions, arising from SSB with type-B GMs. In particular, as an extension of Eq.~(\ref{slnf})
to this type of scale-invariant states in $d$ spatial dimensions, we have	
\begin{equation}
S_{\!\!f} (L^d_g,n^d_g)=\frac {N_B}{ 2 } \log_2 \left(n_g^d (1-  n_g^d /L_g^d)\right ) +S_{\!\!f0}.
\end{equation}
Here $n^d_g$ is defined as $n^d_g=n_1n_2 \ldots n_d$. If we set $L_\delta=L$ and $n_\delta=n$ ($\delta = 1,\ldots,d$), then the entanglement entropy $S_{\!\!f} (L_g,n_g)$ becomes  $S_{\!\!f}(n)$ in the thermodynamic limit. It takes the form
\begin{equation}
S_{\!\!f} (n)= \frac {d\, N_B}{ 2 } \log_2n +S_{\!\!f0}.
\end{equation}
Physically, this results from an observation that the orthonormal basis states admit an exact Schmidt decomposition so that they exhibit self-similarities in the real space, thus revealing an intrinsic abstract fractal underlying the ground state subspace, as already discussed for quantum many-body systems undergoing SSB with type-B GMs in one spatial dimension~\cite{FMGM,cantorset}. As usual, such an intrinsic abstract fractal is characterized by the fractal dimension $d_f$, which in turn is exposed by introducing an extrinsic fractal such as the Cantor sets~\cite{cantorset}.
Here we stress that the fractal dimension $d_f$ has been exploited by Castro-Alvaredo and Doyon~\cite{doyon} for the spin-$1/2$ ${\rm SU}(2)$ ferromagnetic Heisenberg model in one spatial dimension.
We thus conclude that the fractal dimension $d_f$ is identical to the number of type-B GMs, $N_B$, irrespective of the spatial dimensionalities, if only the orthonormal basis states are considered.  It follows that as far as the orthonormal basis states are concerned, the above entanglement entropy scaling relation also takes the equivalent form
\begin{equation}
S_{\!\!f} (n)= \frac {d\, d_f}{ 2 } \log_2n +S_{\!\!f0}.
\end{equation}

\section{Three illustrative examples}~\label{threeexamples}

Our illustrative examples only concern quantum many-body systems on a square lattice, undergoing SSB with type-B GMs. However, our generic scheme also works for other types of two-dimensional lattices. Further,  our prediction makes sense for a ladder or a tube,
given the thermodynamic limit may be achieved if $L_1 \rightarrow \infty$ for fixed $L_2$ and vice versa. This marks a further difference between type-A  and type-B GMs, since {\it only} type-B GMs survive in a quasi-one-dimensional ladder or tube.

\subsection{The ${\rm SU}(2)$ spin-$s$ ferromagnetic Heisenberg model}

The ${\rm SU}(2)$ spin-$s$ ferromagnetic Heisenberg model on a square lattice is described by nearest-neighbor ferromagnetic interactions, with the Hamiltonian
\begin{equation}
\mathscr{H}=-\sum_{\{p,q\}}\textbf{S}_p\cdot \textbf{S}_q, \label{su2ham}
\end{equation}
where $\textbf{S}_p=(S^x_p,S^y_p,S^z_p)$, with $S^x_p$, $S^y_p$, $S^z_p$ denoting the spin-$s$ operators acting at the $p$-th lattice site and the sum over ${\{p,q\}}$ is taken for
the four nearest-neighbor sites on a square lattice.
The highly degenerate ground states arise from SSB from ${\rm SU}(2)$ to ${\rm U}(1)$, with one type-B GM. Hence we have $N_B = 1$.
We remark that the highly degenerate ground states are identical under both OBCs and PBCs, which span an irreducible representation for the symmetry group ${\rm SU}(2)$.

If the highest weight state $|{\rm hws}\rangle$ is chosen to be $|{\rm hws}\rangle=|\otimes_{l_1l_2}\{s\}_{l_1l_2}\rangle$, where $|s\rangle$ is the eigenvector of $S^z_p$ at the $p$-th lattice site, with the eigenvalue being $s$, then highly degenerate ground states $|L_1L_2,M\rangle$  are generated from the repeated action of the lowering operator $S_{-}$, defined as $S_{-}=\sum_p(S_{x,p}- iS_{y,p})/\sqrt{2}$, on the highest weight state $|{\rm hws}\rangle$:
\begin{align}
|L_1L_2,M\rangle=\frac{1}{Z(L,M)}(S_-)^{M}|{\rm hws}\rangle.
\end{align}
Here $Z(L_1L_2,M)$ has been introduced to ensure that $|L_1L_2,M\rangle$ is normalized.  Hence the degenerate ground states$|L_1L_2,M\rangle$ form the orthonormal basis states in the ground state subspace.
A notable observation is that $|L_1L_2,M\rangle$ admits an exact Schmidt decomposition:
\begin{equation}
	|L_1L_2,M\rangle=\sum_{k}\lambda(L_1L_2,k,M) \, |n_1n_2,k\rangle \, |L_1L_2-n_1n_2,M-k\rangle,
	\label{psignln}
\end{equation}
where  $\lambda(L_1L_2,k,M)$ denotes the Schmidt coefficients, which take the form
\begin{equation}
	\lambda(L_1L_2,k,M)=\prod_{\alpha=1}^r\!C_{M}^{k} \frac{Z(n_1n_2,k) \, Z(L_1L_2-n_1n_2,M-k)}{Z(L_1L_2,M)}.
	\label{lambdak}
\end{equation}
Here $C_{M}^{k}$  denotes the binomial coefficient and $Z(n_1n_2,k)$ and $Z(L_1L_2-n_1n_2,M-k)$ are introduced to ensure that $|n_1n_2,k\rangle$ and $|L_1L_2-n_1n_2,M-k\rangle$ are normalized.
As a consequence, the eigenvalues of the reduced density matrix $\rho(L_1L_2,n_1n_2)$  follow from the Schmidt coefficients, as collected in Sec. C of the SM. This in turn allows us to evaluate the entanglement entropy $S(L_1L_2,n_1n_2,M)$.

\subsection{The ${\rm SU}(2s+1)$ ferromagnetic model}

The ${\rm SU}(2s+1)$ ferromagnetic model on a square lattice  is described by the Hamiltonian
\begin{equation}
\mathscr{H}=-\sum_{\{p,q\}} P_{\{p,q\}}, \label{HsuNp1}
\end{equation}
where $P$ is the permutation operator and the sum over ${\{p,q\}}$ is taken for
the four nearest-neighbor sites on a square lattice.
Physically, the permutation operator $P$ may be realized in terms of the spin-$s$ operators $\textbf{S}=(S_x,S_y,S_z)$.
Indeed, when $s=1$, the model constitutes a special point of the ${\rm SU}(3)$ spin-1  bilinear-biquadratic model  on a square lattice.
We remark that the ${\rm SU}(2s+1)$ ferromagnetic model on a square lattice is exactly solvable, as far as the ground state subspace is concerned. However, this is not true if one attempts to go beyond the ground state subspace, in contrast to its one-dimensional version, which is exactly solvable by means of the nested Bethe ansatz~\cite{sutherland}.
The highly degenerate ground states arise from SSB from ${\rm SU}(2s+1)$ to ${\rm SU}(2s) \times {\rm U}(1)$ successively, with $2s$ type-B GMs. Hence we have $N_B = 2s$.
Note that the highly degenerate ground states are identical under both OBCs and PBCs, and span an irreducible representation for the symmetry group ${\rm SU}(2s+1)$, with the dimension being the binomial coefficient $C_{L_1L_2+2s}^{2s}$. They thus form the orthonormal basis states in the ground state subspace.

If the highest weight state $|\rm{hws}\rangle$ is chosen to be $|{\rm hws}\rangle=|\otimes_{l_1l_2}\{s\}_{l_1l_2}\rangle$, where $|s\rangle$ is the eigenvector of $S^z_p$ at the $p$-th lattice site, with the eigenvalue being $s$,  then highly degenerate ground states $|L_1L_2,M_1,\ldots,M_{2s}\rangle$ are generated from the repeated action of the lowering operators $F_{\alpha}$ ($\alpha=1$,\ldots, $2s$) on the highest weight state $|{\rm hws}\rangle$:
\begin{equation}
|L_1L_2,M_1,\ldots,M_{2s}\rangle=\frac{1}{Z(L_1L_2,M_1,\ldots,M_{2s})}\prod_{\alpha=1}^{2s}F_\alpha^{\,\,M_\alpha}|{\rm hws}\rangle,	\label{lm1mr}
\end{equation}
where $Z(L_1L_2,M_1,\ldots,M_{2s})$ is introduced to ensure that $|L_1L_2,M_1,\ldots,M_{2s}\rangle$ is normalized.

The degenerate ground state $|L_1L_2,M_1,\ldots,M_{2s}\rangle$ in Eq.~(\ref{lm1mr}) admits an exact Schmidt decomposition:
\begin{equation}
	|L_1L_2,M_1,\ldots,M_{2s}\rangle=\prod_{\alpha=1}^{2s}\sum_{k_\alpha}\lambda(L_1L_2,k_1,\ldots,k_{2s},M_1,\ldots,M_{2s}) \, 
	|n_1n_2,k_1,\ldots,k_{2s}\rangle \, |L_1L_2-n_1n_2,M_1-k_1,\ldots,M_{2s}-k_{2s}\rangle,
	\label{psignln}
\end{equation}
where  $\lambda(L_1L_2,k_1,\ldots,k_{2s},M_1,\ldots,M_{2s})$ denote the Schmidt coefficients, which take the form
\begin{equation}
	\lambda(L_1L_2,k_1,\ldots,k_{2s},M_1,\ldots,M_{2s})=\prod_{\alpha=1}^{2s}C_{M_\alpha}^{k_\alpha} 
	\frac{Z(n_1n_2,k_1,\ldots,k_{2s}) \, Z(L_1L_2-n_1n_2,M_1-k_1,\ldots,M_{2s}-k_{2s})}{Z(L_1L_2,M_1,\ldots,M_{2s})}.
	\label{lambdak}
\end{equation}
Here $C_{M_\alpha}^{k_\alpha}$  denotes the binomial coefficient and $Z(n_1n_2,k_1,\ldots,k_{2s})$ and $Z(L_1L_2-n_1n_2,M_1-k_1,\ldots,M_{2s}-k_{2s})$ are introduced to ensure that $|n_1n_2,k_1,\ldots,k_{2s}\rangle$ and $|L_1L_2-n_1n_2,M_1-k_1,\ldots,M_{2s}-k_{2s}\rangle$ are normalized.
As a consequence, the eigenvalues of the reduced density matrix $\rho(L_1L_2,n_1n_2)$ follow from the Schmidt coefficients, as collected in Sec. C of the SM. This in turn allows us to evaluate the entanglement entropy $S(L_1L_2,n_1n_2,M_1,\ldots,M_{2s})$.

\subsection{The staggered ${\rm SU}(3)$ ferromagnetic biquadratic model}

The staggered ${\rm SU}(3)$ spin-1 ferromagnetic biquadratic model on a square lattice is described by the Hamiltonian
\begin{equation}
\mathscr{H}=\sum_{\{p,q\}}{\left(\textbf{S}_p \cdot \textbf{S}_q \right)^2}, \label{hambq}
\end{equation}
where $\textbf{S}_p=(S^x_p,S^y_p,S^z_p)$, with $S^x_p$, $S^y_p$, $S^z_p$ denoting the spin-$1$ operators acting at the $p$-th lattice site and the sum over ${\{p,q\}}$ is taken for
the four nearest-neighbor sites on a square lattice.
The model (\ref{hambq}) is also exactly solvable, as far as the ground state subspace is concerned. However, this is not true if one attempts to go beyond the ground state subspace, in contrast to its one-dimensional version, which is exactly solvable  in the context of the representations of the Temperley-Lieb algebra or the quantum Yang-Baxter equation~\cite{barber}.
Note that the Hamiltonian (\ref{hambq}) is a special point in the spin-1 bilinear-biquadratic model on a square lattice~\cite{bb2d}.
For this model the highly degenerate ground states arise from SSB from ${\rm SU}(3)$ to ${\rm U}(1) \times {\rm U}(1)$, with the number of the type-B GMs $N_B=2$.
We emphasize that  the ground state degeneracies  under both OBCs and PBCs are exponential with the system size $L_1L_2$, thus leading to a non-zero residual entropy, similar to the one-dimensional version~\cite{klumper}. However, it remains challenging to figure out exactly the ground state degeneracies under both OBCs and PBCs.

The model (\ref{hambq}) admits both the highest weight state $|{\rm hws}\rangle$  and generalized highest weight states $|{\rm hws}\rangle^g_{p_1p_2}$, where  $p_1$ and  $p_2$ are the periods in the horizontal and vertical directions, respectively. The highest weight state $|{\rm hws}\rangle$ is chosen to be $|{\rm hws}\rangle=|\otimes_{l_1l_2}\{1\}_{l_1l_2}\rangle$, where  $|1\rangle$ is the eigenvector of spin-1 operator $S_{z,j}$, with the eigenvalue being $1$. Generically, generalized highest weight state $|{\rm hws}\rangle^g_{p_1p_2}$ may be represented as  $|\otimes_{l_l,l_2}\{s_{11}s_{21}...s_{p_1p_2}\}_{l_l,l_2}\rangle$, where $l_1$ ranges from $1$ to $L_1/p_1$ and $l_2$ ranges from $1$ to $L_2/p_2$. Note that $L_1$ is a multiple of $p_1$ and $L_2$ is a multiple of $p_2$. In particular,  we focus on a specific generalized highest weight state  $|{\rm hws}\rangle^g_{44}=|\otimes_{l_1l_2}\{1110110110110111\}_{l_1l_2}\rangle$, with the periods $p_1$ and $p_2$ being $4$.
Highly degenerate ground states, which appear as the orthonormal basis states in a given sector of the ground state subspace, are generated from the repeated action of the lowering operators $F_1$ and $F_2$~\cite{FMGM} on either the  highest weight state $|{\rm hws}\rangle$  or a chosen generalized highest weight state $|{\rm hws}\rangle^g_{p_1p_2}$. The latter yields the highly degenerate ground states with the periods $q_1$ and $q_2$ in the horizontal and vertical directions, denoted as $|L_1L_2,M_1,M_2\rangle_{q_1q_2}$.
For our purpose here, we focus on two sequences of highly degenerate ground states, namely $|L_1L_2,M_1,M_2\rangle_{22}$, with $q_1=q_2=2$, and the highly degenerate ground states $|L_1L_2,M_2,M_3\rangle_{44}$, with $q_1=q_2=4$.

Highly degenerate ground states $|L_1L_2,M_1,M_2\rangle_{22}$, with $q_1=q_2=2$,  are generated from the repeated action of the lowering operators $F_1$ and $F_2$ on the highest weight state $|{\rm hws}\rangle=|\otimes_{l_1l_2}\{1\}_{l_1l_2}\rangle$:
\begin{equation}
|L_1L_2,M_1,M_2\rangle_{22}=\frac{1}{Z_{22}(L_1L_2,M_1,M_2)}F_1^{M_1}F_2^{M_2}|{\rm hws}\rangle,
\label{lm1m2}
\end{equation}
where  $Z_{22}(L_1L_2,M_1,M_2)$ is introduced to ensure that $|L_1L_2,M_1,M_2\rangle_{22}$ is normalized.

The degenerate ground state $|L_1L_2,M_1,M_2\rangle_{22}$ in Eq.~(\ref{lm1m2}) admits an exact Schmidt decomposition:
\begin{equation}
	|L_1L_2,M_1,M_2\rangle_{22}=\sum_{k_1,k_2}\lambda(L_1L_2,k_1,k_2,M_1,M_2) \, 
	|n_1n_2,k_1,k_2\rangle_{22} \, |L_1L_2-n_1n_2,M_1-k_1,M_2-k_2\rangle_{22},
	\label{psignln}
\end{equation}
where  $\lambda(L_1L_2,k_1,k_2,M_1,M_2)$ denote the Schmidt coefficients, which take the form
\begin{equation}
	\lambda(L_1L_2,k_1,k_2,M_1,M_2)=C_{M_1}^{k_1}C_{M_2}^{k_2} 
	\frac{Z_{22}(n_1n_2,k_1,k_2)Z_{22} \, (L_1L_2-n_1n_2,M_1-k_1,M_2-k_2)}{Z_{22}(L_1L_2,M_1,M_2)}.
	\label{lambdak}
\end{equation}
Here $Z_{22}(n_1n_2,k_1,k_2)$ and $Z_{22}(L_1L_2-n_1n_2,M_1-k_1,M_2-k_2)$ are introduced to ensure that $|n_1n_2,k_1,k_2\rangle$ and $|L_1L_2-n_1n_2,M_1-k_1,M_2-k_2\rangle_{22}$ are normalized.
As a consequence, the eigenvalues of the reduced density matrix $\rho(L_1L_2,n_1n_2)$  follow from the Schmidt coefficients, as collected in Sec.~C of the SM. This in turn allows us to evaluate the entanglement entropy $S(L_1L_2,n_1n_2,M_1,M_2)$.

Highly degenerate ground states $|L_1L_2,M_2,M_3\rangle_{44}$, with $q_1=q_2=4$, are generated from the repeated action of the lowering operators $F_2$ and $F_3$ on the generalized highest weight state $|{\rm hws}\rangle^g_{44}$
\begin{equation}
	|L_1L_2,M_2,M_3\rangle_{44}=\frac{1}{Z_{44}(L,M_2,M_3)}F_2^{M_2}F_3^{M_3}|{\rm hws}\rangle^g_{44},
	\label{lm2m3}
\end{equation}
where $Z_{44}(L_1L_2,M_2,M_3)$ is introduced to ensure that $|L,M_2,M_3\rangle_{44}$ is normalized.

The degenerate ground state $|L_1L_2,M_2,M_3\rangle_{44}$ in Eq.~(\ref{lm2m3}) admits an exact Schmidt decomposition:
\begin{equation}
	|L_1L_2,M_2,M_3\rangle_{44}=\sum_{k_2,k_3}\lambda(L_1L_2,k_2,k_3,M_2,M_3) \, 
	|n_1n_2,k_2,k_3\rangle_{44} \, |L_1L_2-n_1n_2,M_2-k_2,M_3-k_3\rangle_{44},
	\label{psignln}
\end{equation}
where  $\lambda(L_1L_2,k_2,k_3,M_2,M_3)$ denote the Schmidt coefficients, which take the form
\begin{equation}
	\lambda(L_1L_2,k_1,k_2,M_2,M_3)=C_{M_2}^{k_2}C_{M_3}^{k_3} \, 
	\frac{Z_{44}(n_1n_2,k_2,k_3)Z_{44} \, (L_1L_2-n_1n_2,M_2-k_2,M_3-k_3)}{Z_{44}(L_1L_2,M_2,M_3)}.
	\label{lambdak}
\end{equation}
Here $Z_{44}(n_1n_2,k_2,k_3)$ and $Z_{44}(L_1L_2-n_1n_2,M_2-k_2,M_3-k_3)$ are introduced to ensure that $|n_1n_2,k_2,k_3\rangle$ and $|L_1L_2-n_1n_2,M_2-k_2,M_3-k_3\rangle_{44}$ are normalized.
As a consequence, the eigenvalues of the reduced density matrix $\rho(L_1L_2,n_1n_2)$  follow from the Schmidt coefficients, as collected in Sec.~C of the SM. This in turn allows us to evaluate the entanglement entropy $S(L_1L_2,n_1n_2,M_2,M_3)$.
Here $|L_1L_2,M_2,M_3\rangle_{44}$ is denoted as $|L_1L_2,M_1^*,M_2^*\rangle_{44}$ and $S(L_1L_2,n_1n_2,M_2,M_3)$ is denoted as $S(L_1L_2,n_1n_2,M_1^*,M_2^*)$, where the fillings  $f_1^*$ and  $f_2^*$ are defined as $f_1^*=M_1^*/L$ and $f_2^*=M_2^*/L$, with $M_1^*=M_2$ and $M_2^*=M_3$.

\section{Entanglement entropy for scale-invariant states in two spatial dimensions}~\label{ee}

Now we perform a finite system-size scaling analysis of the entanglement entropy to numerically confirm the theoretical prediction (\ref{slnf}). Our numerics are carried out for the orthonormal basis states in the ${\rm SU}(2)$ spin-$s$ ferromagnetic Heisenberg model, the ${\rm SU}(2s+1)$ ferromagnetic model, and the staggered ${\rm SU}(3)$ spin-1 ferromagnetic biquadratic model.

\begin{figure}[htb]
	\centering
	\includegraphics[width=0.4\textwidth]{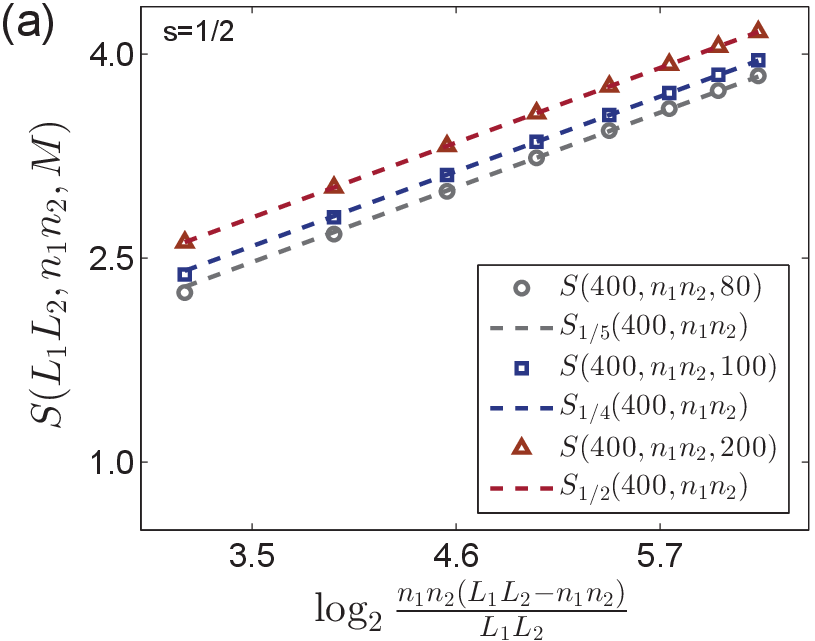}~~~~
	\includegraphics[width=0.4\textwidth]{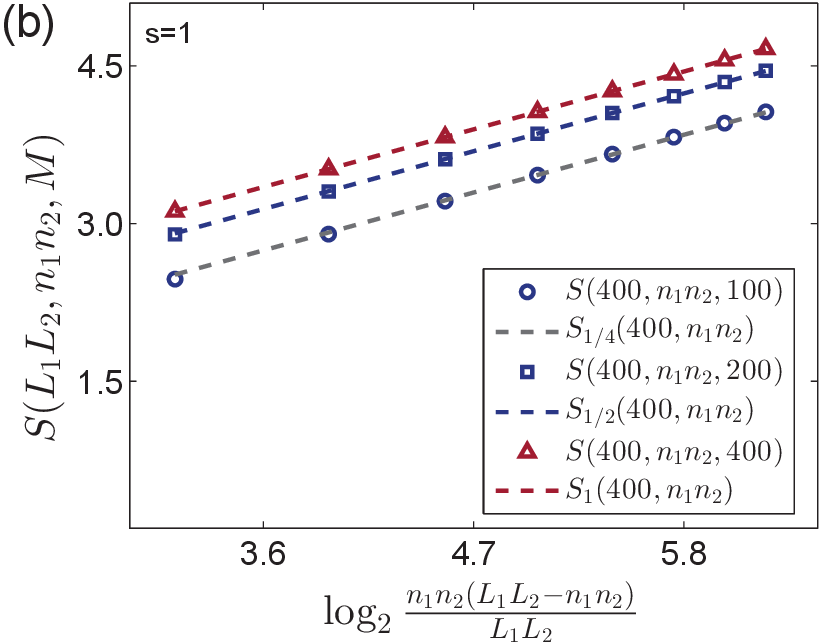}
	\caption{Entanglement entropy scaling for the highly degenerate ground states $|L_1L_2,M\rangle$  of the ${\rm SU}(2)$ spin-$s$ ferromagnetic Heisenberg model for (a) $s=1/2$ and (b) $s=1$ on the square lattice of size $L_1=L_2=20$. 
The entanglement entropy $S(L_1L_2,n_1n_2,M)$ is plotted as a function of $\log_2(n_1n_2(L_1L_2-n_1n_2)/(L_1L_2))$, with $S_{\!\!f}(L_1L_2,n_1n_2)$ vs $\log_2(n_1n_2(L_1L_2-n_1n_2)/(L_1L_2))$ shown for comparison. Best fitting yields (a) $S_{\!\!f0}=0.7211$, $0.838$ and $1.049$ when the filling $f$ varies as $f=1/5$, $1/4$ and $1/2$; (b)  $S_{\!\!f0}= 0.945$, $1.339$ and $1.548$  when the filling $f$ varies as $f=1/4$, $1/2$ and $1$. The relative errors are less than $2\%$, as $n_1$ and $n_2$ range from 3 to 10.}
	\label{comparesu2spins}
\end{figure}

Fig.~\ref{comparesu2spins} shows the entanglement entropy $S(L_1L_2,n_1n_2,M)$ plotted  as a function of $\log_2(n_1n_2(L_1L_2-n_1n_2)/(L_1L_2))$ 
for the two-dimensional ${\rm SU}(2)$ spin-$s$ ferromagnetic Heisenberg model, with $s=1/2$ and $s=1$, when the filling $f$ varies. 
The highest weight state is chosen to be $|{\rm hws}\rangle=|\otimes_{l_1l_2}\{s\}_{l_1l_2}\rangle$, where $|s\rangle$ is the eigenvector of $S^z_p$ at the $p$-th lattice site, with the eigenvalue being $s$. 
The results are compared against $S_{\!\!f}(L_1L_2,n_1n_2)$ vs $\log_2(n_1n_2(L_1L_2-n_1n_2)/(L_1L_2))$.
The numerical data for $S(L_1L_2,n_1n_2,M)$ is seen to fall on a straight line $S_{\!\!f}(L_1L_2,n_1n_2)$ vs $\log_2(n_1n_2(L_1L_2-n_1n_2)/(L_1L_2))$ according to the predicted scaling relation (\ref{slnf}).
Here and hereafter, we have regarded $S_{\!\!f}(L_1L_2,n_1n_2)$ as a function of $n_1n_2$ for fixed $L_1L_2$ and $f$.

\begin{figure}[htb]
	\centering
	\includegraphics[width=0.4\textwidth]{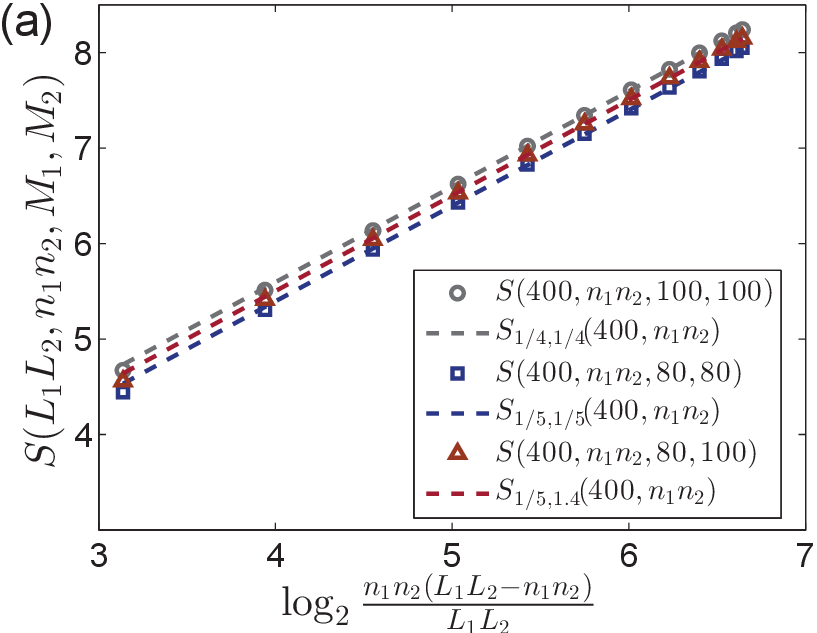}~~~~
	\includegraphics[width=0.4\textwidth]{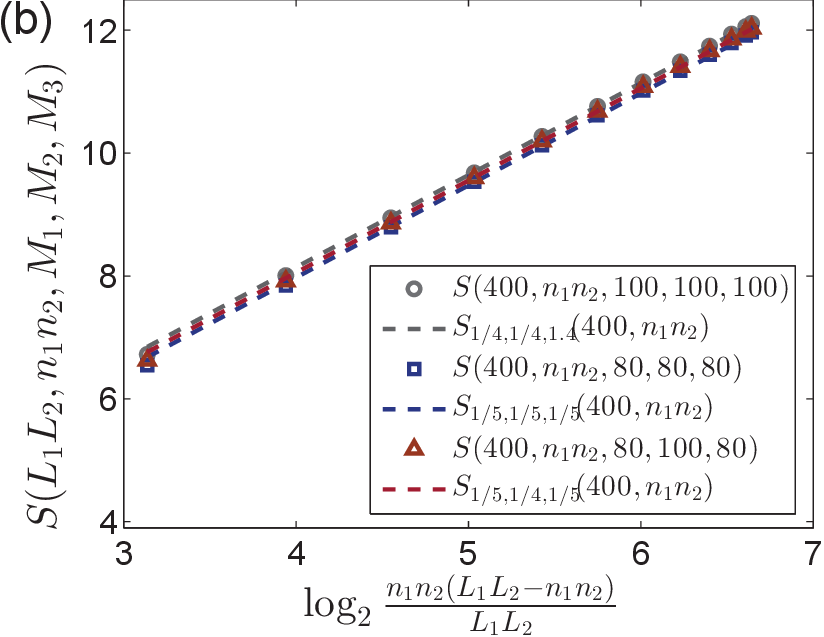}
	\caption{Entanglement entropy scaling for the highly degenerate ground states of the ${\rm SU}(2s+1)$ ferromagnetic model on the square lattice, with spin (a) $s=1$ and (b) $s=3/2$. 
	(a) The entanglement entropy $S(L_1L_2,n_1n_2,M_1,M_2)$ vs $\log_2(n_1n_2(L_1L_2-n_1n_2)/(L_1L_2))$ for the highly degenerate ground states $|L_1L_2,M_1,M_2\rangle$  of the ${\rm SU}(3)$ spin-$1$ model.
		(b) The entanglement entropy $S(L_1L_2,n_1n_2,M_1,M_2,M_3)$ vs $\log_2(n_1n_2(L_1L_2-n_1n_2)/(L_1L_2))$ for the highly degenerate ground states $|L_1L_2,M_1,M_2,M_3\rangle$  of the ${\rm SU}(4)$ spin-$3/2$ model.
For both cases the plot of $S_{\!\!f}(L_1L_2,n_1n_2)$ vs $\log_2(n_1n_2(L_1L_2-n_1n_2)/(L_1L_2))$ is shown for comparison. 
		Here $L_1=L_2=20$, with $n_1$ and $n_2$ ranging from 3 to 14.
		Best fitting yields (a) $S_{\!\!f0}=1.595$, $1.395$ and $1.502$ for the fillings $(f_1,f_2)=(1/4,1/4)$, $(1/5,1/5)$, and $(1/5,1/4)$; (b) $S_{\!\!f0}=2.141$, $1.975$ and $2.055$ for the fillings $(f_1,f_2,f_3)=(1/4,1/4,1/4)$, $(1/5,1/5,1/5)$, and $(1/5,1/4,1/5)$.  The relative errors are less than $2\%$.}
	\label{comparesu3su4}
\end{figure}

Fig.~\ref{comparesu3su4} shows similar plots for the ${\rm SU}(2s+1)$ ferromagnetic model on the square lattice, with spin $s=1$ and $s=3/2$.
For each case the entanglement entropy $S(L_1L_2,n_1n_2,M_1,M_2)$ and the entanglement entropy $S(L_1L_2,n_1n_2,M_1,M_2,M_3)$ are plotted as a function of 
$\log_2(n_1n_2(L_1L_2-n_1n_2)/(L_1L_2))$, when the fillings $f_1$ and $f_2$ for $s=1$ and the fillings $f_1$, $f_2$ and $f_3$ for $s=3/2$ vary. 
Here, the highest weight state is chosen to be $|{\rm hws}\rangle=|\otimes_{l_1l_2}\{s\}_{l_1l_2}\rangle$, where $|s\rangle$ is the eigenvector of $S^z_p$ at the $p$-th lattice site, with the eigenvalue being $s$.
Shown also for comparison is a plot of $S_{\!\!f}(L_1L_2,n_1n_2)$ vs $\log_2(n_1n_2(L_1L_2-n_1n_2)/(L_1L_2))$, according to the scaling relation (\ref{slnf}), with $N_B=2$ and $N_B=3$.
The numerical data for $S(L_1L_2,n_1n_2,M_1,M_2)$ and $S(L_1L_2,n_1n_2,M_1,M_2,M_3)$ is clearly seen to fall on the predicted line.

\begin{figure}[htb]
	\centering
	\includegraphics[width=0.4\textwidth]{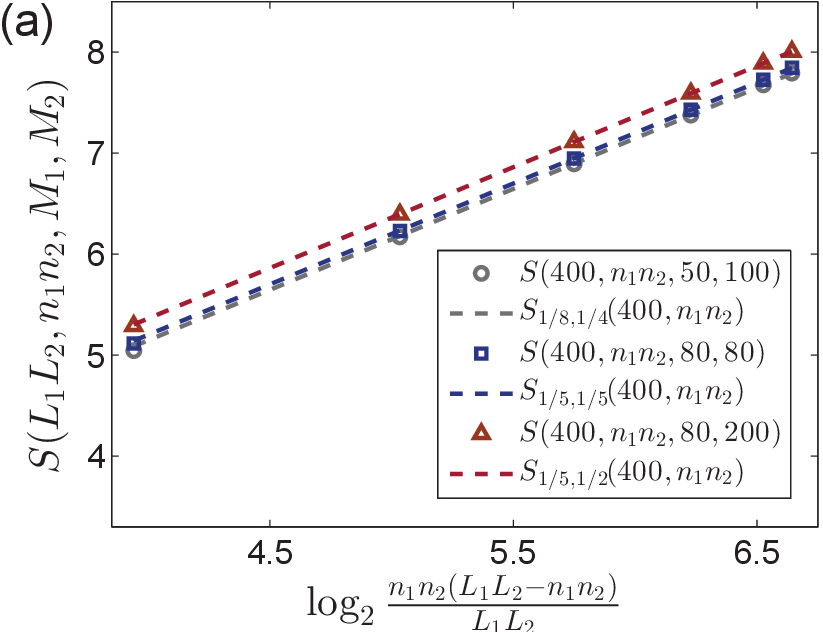}~~~~
	\includegraphics[width=0.4\textwidth]{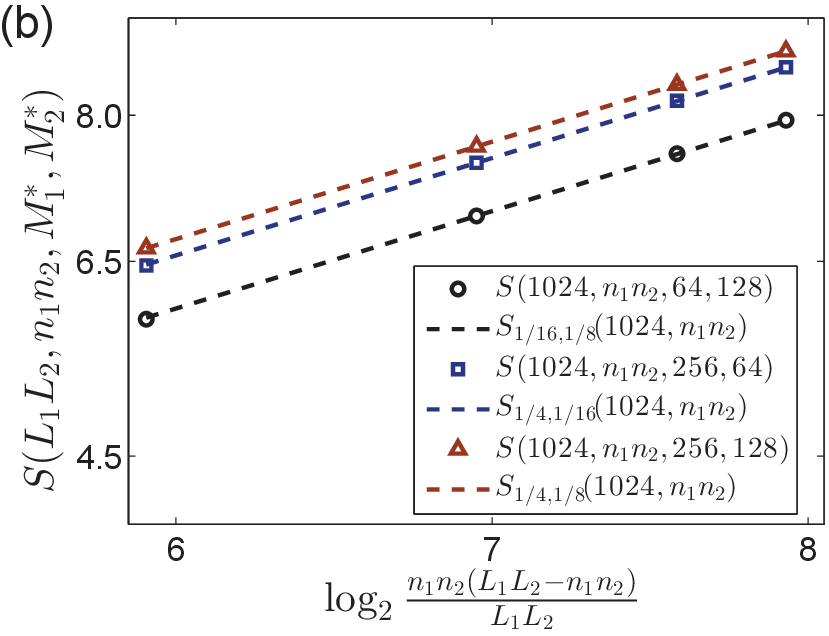}
	\caption{Entanglement entropy scaling for the highly degenerate ground states of the staggered ${\rm SU}(3)$ spin-1 ferromagnetic biquadratic model on the square lattice.
	(a) The entanglement entropy $S(L_1L_2,n_1n_2,M_1,M_2)$  vs $\log_2(n_1n_2(L_1L_2-n_1n_2)/(L_1L_2))$ for the highly degenerate ground states  $|L_1L_2,M_1,M_2\rangle_{22}$  from the highest weight state, with $L_1=L_2=20$, when the fillings $f_1$ and $f_2$ vary. Here, the highest weight state is chosen to be $|{\rm hws}\rangle=|\otimes_{l_1l_2}\{1\}_{l_1l_2}\rangle$, where $|1\rangle$ is the eigenvector of $S^z_p$ at the $p$-th lattice site, with the eigenvalue being $1$.
	Best fitting yields $S_{\!\!f0}=1.146$, $1.200$ and $1.362$ for the fillings $(f_1,f_2)=(1/8,1/4)$, $(1/5,1/5)$ and $(1/5,1/2)$, with the relative errors less than $1\%$, as $n_1$ and $n_2$ range from 4 to 14.
	(b) The entanglement entropy $S(L_1L_2,n_1n_2,M_1^*,M_2^*)$ vs $\log_2(n_1n_2(L_1L_2-n_1n_2)/(L_1L_2))$ for the highly degenerate ground states $|L_1L_2,M_1^*,M_2^*\rangle_{44}$ from a generalized highest weight state, with $L_1=L_2=32$, when the fillings $f_1^*$ and $f_2^*$ vary.  Here the generalized highest weight state is chosen to be $|{\rm hws}\rangle^g_{44}=|\otimes_{l_1l_2}\{1110110110110111\}_{l_1l_2}\rangle$ in a periods of 4, i.e., $p_1=p_2=4$, where $|1\rangle$ and $|0\rangle$ are the eigenvectors of $S^z_p$ at the $p$-th lattice site, with the eigenvalues being $1$ and $0$, respectively.
	Best fitting yields $S_{\!\!f0}=0.0160$, $0.561$ and $0.729$ for the fillings $(f_1^*,f_2^*)=(1/16,1/8)$, $(1/4,1/16)$ and $(1/4,1/8)$, with the relative errors less than $1\%$, as $n_1$ and $n_2$ range from 8 to 20.}
	\label{comparestsu3}
\end{figure}

In Fig.~\ref{comparestsu3}, we plot the entanglement entropy for the staggered ${\rm SU}(3)$ spin-1 ferromagnetic biquadratic model on the square lattice. 
Specifically the entanglement entropy $S(L_1L_2,n_1n_2,M_1,M_2)$ and $S(L_1L_2,n_1n_2,M_2,M_3)$ as a function of $\log_2(n_1n_2(L_1L_2-n_1n_2)/(L_1L_2))$. 
Fig.~\ref{comparestsu3} (a) corresponds to the highest weight state $|{\rm hws}\rangle=|\otimes_{l_1l_2}\{1\}_{l_1l_2}\rangle$, where $|1\rangle$ is the eigenvector of $S^z_p$ at the $p$-th lattice site, with the eigenvalue being $1$. Fig.~\ref{comparestsu3} (b) corresponds to the generalized highest weight state $|{\rm hws}\rangle^g_{44}=|\otimes_{l_1l_2}\{1110110110110111\}_{l_1l_2}\rangle$  with $p_1=p_2=4$, where $|1\rangle$ and $|0\rangle$ are the eigenvectors of $S^z_p$ at the $p$-th lattice site, with the eigenvalues being $1$ and $0$, respectively.
Here $S(L_1L_2,n_1n_2,M_2,M_3)$ is denoted as $S(L_1L_2,n_1n_2,M_1^*,M_2^*)$, where the fillings  $f_1^*$ and  $f_2^*$ are defined as $f_1^*=M_1^*/L$ and $f_2^*=M_2^*/L$, with $M_1^*=M_2$ and $M_2^*=M_3$.
The numerical data for $S(L_1L_2,n_1n_2,M_1,M_2)$ and $S(L_1L_2,n_1n_2,M_1^*,M_2^*)$ are again seen to fall on the straight line defined by $S_{\!\!f}(L_1L_2,n_1n_2)$ vs  $\log_2(n_1n_2(L_1L_2-n_1n_2)/(L_1L_2))$ when the fillings $f_1$ and $f_2$ or the fillings $f_1^*$ and $f_2^*$ vary, according to the scaling relation (\ref{slnf}).

\section{Summary}~\label{summary}

We have presented a generic scheme to perform a universal finite system-size scaling analysis for the entanglement entropy in quantum many-body systems undergoing SSB with  type-B GMs, valid in two spatial dimensions and beyond, as far as the orthonormal basis states are concerned. It heavily relies on an observation that highly degenerate ground states arising from SSB with  type-B GMs are scale-invariant, thus imposing the three physical constraints on the universal logarithmic scaling term representing the dominant contribution to the entanglement entropy, if these degenerate ground states are permutation-invariant. A notable feature is that the contribution from the area law to the entanglement entropy is absent for almost all known quantum many-body systems exhibiting SSB with type-B GMs, since both the closeness of degrees of freedom to the boundary between a subsystem and its environment and the boundary itself are  not well-defined, depending on the type of a permutation operation performed. An abstract fractal underlying the ground state subspace, characterized by the fractal dimension, manifests itself in the exact Schmidt decomposition for the orthonormal basis states, thus leading to the identification of the fractal dimension with  the number of type-B GMs for the orthonormal basis states.

In addition, a few remarkable differences between type-A and type-B GMs have been revealed. In particular, quantum many-body systems undergoing SSB with type-B GMs constitute a realization of the simplest spatial fine structure quantified by the entanglement contour~\cite{guifre}, if degenerate ground states arising from SSB with type-B GMs are permutation-invariant.
In a sense, they are {\it trivial} from a perspective of  the entanglement contour, in sharp contrast to those undergoing SSB with type-A GMs in two and higher spatial dimensions as well as critical quantum many-body systems described by conformal field theory in one spatial dimension. However,  the  flat-band ferromagnetic Tasaki model is an exception, in the sense that degenerate ground states are not always permutation-invariant~\cite{TypeBtasaki}.
The entanglement entropy for the two-dimensional flat-band ferromagnetic Tasaki model therefore deserves further investigation.

The predicted entanglement entropy scaling relation (\ref{slnf}) has been tested for three fundamental quantum many-body systems on the square lattice -- the ${\rm SU}(2)$ spin-$s$ ferromagnetic Heisenberg model, the ${\rm SU}(2s+1)$ ferromagnetic model, and the staggered ${\rm SU}(3)$ spin-1 ferromagnetic biquadratic model.
One may anticipate that our discussions are also applicable to other lattice types in two spatial dimensions and beyond.

\section{Acknowledgements}

We thank John Fjaerestad  for enlightening discussions and carefully reading the manuscript and Jesse Osborne for helpful discussions on the flat-band ferromagnetic Tasaki model.

\newpage
\onecolumngrid
\newpage
\section*{Supplementary Material}
\setcounter{page}{1}
\setcounter{equation}{0}
\setcounter{figure}{0}
\renewcommand{\theequation}{S\arabic{equation}}
\renewcommand{\thefigure}{S\arabic{figure}}
\renewcommand{\bibnumfmt}[1]{[S#1]}
\renewcommand{\citenumfont}[1]{S#1}

\subsection{Exact MPS representations for degenerate ground states arising from SSB with type-B GMs in two spatial dimensions}
For quantum many-body systems on a lattice in two spatial dimensions, it is conceptually natural to resort to a projected entangled pair state (PEPS) representation~\cite{PEPSsm} for ground state wave functions. However, a remarkable observation is that
it is possible to expose an exact MPS representation  for highly degenerate ground states arising from SSB with type-B GMs in two and higher spatial dimensions, at least for those with  the highest weight state or a generalized highest weight state (if any) being (unentangled) factorized. Physically, this is largely due to the fact that there exists an emergent permutation symmetry group with respect to the unit cells  for highly degenerate ground states arising from SSB with type-B GMs.

\begin{figure}
	\centering
	\includegraphics[angle=0,totalheight=3.5cm]{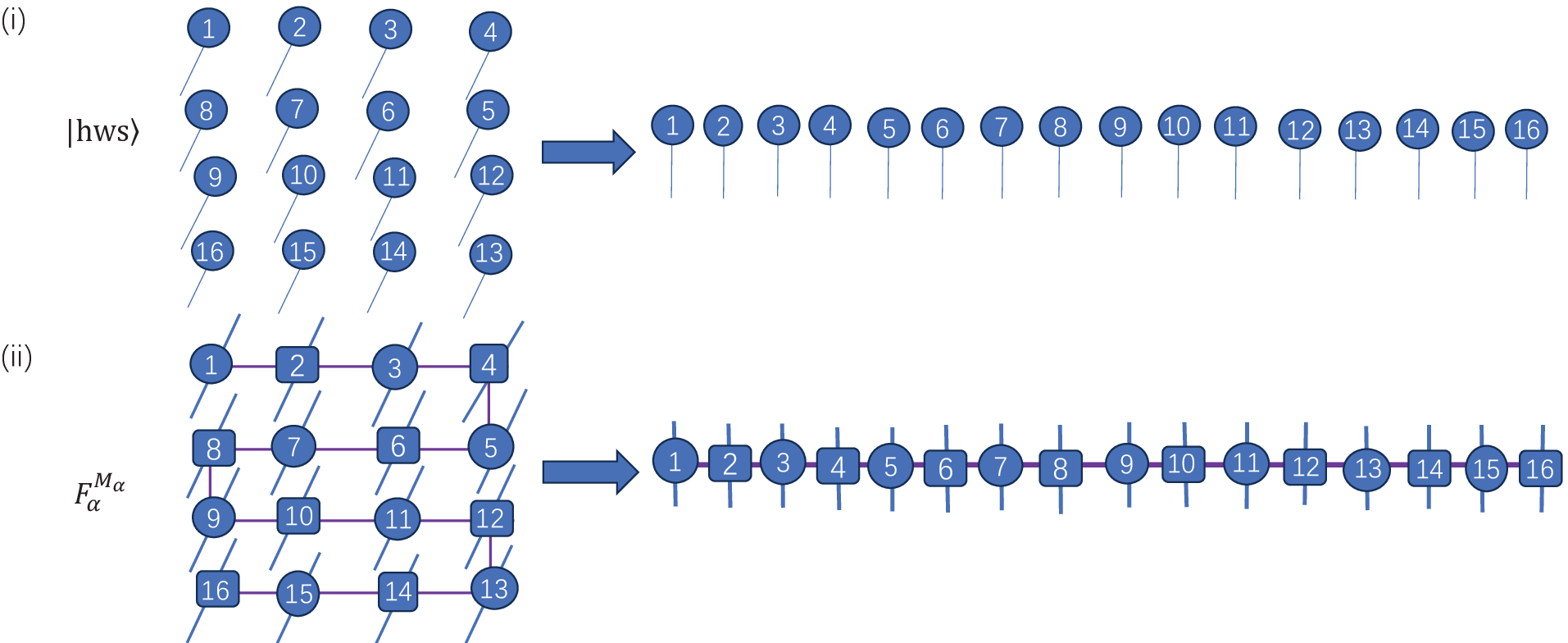}
	\caption{(i) A MPS representation for a factorized ground state $|{\rm hws}\rangle=\otimes_{l=1}^{16}\{s\}_l\rangle$ on a two-dimensional square lattice consisting of 16 lattice sites, irrespective of what types of the boundary conditions, i.e., OBCs and PBCs, are adopted.   All of the lattice sites are reordered, labeled by a number inside the circles.  We are thus able to construct a MPS representation for a factorized ground state, with the bond dimension being $1$.
	(ii) A MPO representation for $F_\alpha^{\;M_\alpha}$ ($\alpha=1$, 2, \ldots, $r$) is constructed, compatible with the staggered nature of the lowering operators, with circles and squares to denote lattice sites on the two sub-lattices, respectively,  for the staggered ${\rm SU}(3)$ spin-1  ferromagnetic biquadratic model.
    A MPO representation for $F_\alpha^{\,\,M_\alpha}$ consists of two vectors at the two ends and the bulk matrices in between. Here the bulk matrices are either uniform or staggered,  depending on the nature of the symmetry group $G$.
	}\label{pepsmps}
\end{figure}

A sequence of highly degenerate ground states are generated from the repeated action of the lowering operators $F_\alpha$ on the highest weight state or a generalized highest weight state. Here, we assume that the highest weight state or a generalized highest weight state is factorized,
as shown in Fig.~\ref{pepsmps} (i), with the highest weight state $|{\rm hws}\rangle$ as an example. After reordering the lattice sites labeled by a number inside the circles, we are capable of deforming a two-dimensional (square) lattice into a one-dimensional chain, irrespective of OBCs or PBCs, as far as degenerate ground states are concerned. Accordingly, the highest weight state admits a MPS representation,  with the bond dimension being $1$.  In addition, we may construct a matrix product operator (MPO) representation for the operator $F_\alpha^{\;M_\alpha}$ ($\alpha=1$, 2, \ldots, $r$), which appears as a power of the lowering operators  $F_\alpha$, as long as all the lattice sites  are ordered according to the deformation described above. In Fig.~\ref{pepsmps} (ii), a MPO representation for $F_\alpha^{\;M_\alpha}$ on a $4\times 4$ square lattice is presented, compatible with the staggered nature of the lowering operators, with  circles and squares to denote lattice sites on the two sub-lattices, for the staggered ${\rm SU}(3)$ spin-1 ferromagnetic biquadratic model on a square lattice. That is, the bulk matrices in the MPO representation are either uniform or staggered, depending on the nature of the symmetry group $G$.  We stress that if a symmetry group is uniform,  then it is not necessary to distinguish circles from squares on the two sub-lattices, as happens for  the ${\rm SU}(2)$ spin-$s$ ferromagnetic Heisenberg model and the ${\rm SU}(2s+1)$ ferromagnetic model.

\begin{figure}
	\centering
	\includegraphics[angle=0,totalheight=3.4cm]{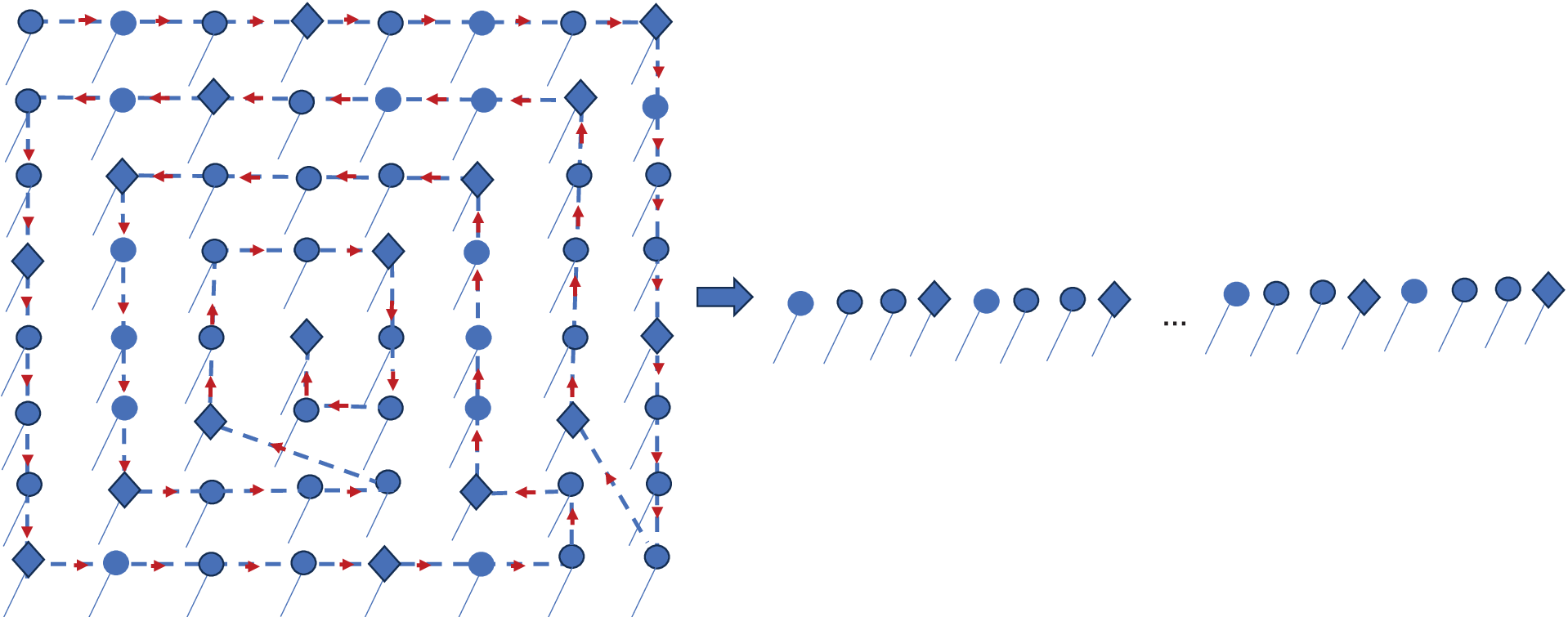}
	\caption{ A MPS representation, with the size of the unit cells being 4,  for a generalized highest weight state, which is a factorized ground state $|{\rm ghws}\rangle_4$, with the period $p_c$ being 4, if the two-dimensional lattice is deformed into a one-dimensional chain.  Here,  circles and diamonds denote the basis states $|1\rangle$ and $|0\rangle$, respectively. Hence, a MPS representation for this generalized highest weight state on a $8\times 8$ square lattice may be written as $|{\rm ghws}\rangle_4=|\otimes_{l=1}^{16}\{1110\}_l\rangle$, where the subscript 4 indicates the period.
	}\label{pepsmpsn4}
\end{figure}

An extension to a generalized highest weight state is straightforward, if it is factorized. Generically,
a generalized highest weight state may be regarded as a PEPS with the bond dimension being $1$, denoted as $|hws\rangle^g_{p_1p_2}$, where $p_1$ and $p_2$ denote the periods in the horizontal and vertical directions, respectively. Such a PEPS representation may be reduced to a MPS representation, with the bond dimension being $1$, denoted as $|ghws\rangle_{p_c}$, where  $p_c$ is the period in a MPS presentation. We emphasize that the period $p_c$ depends {\it not only} on the periods $p_1$ and $p_2$, {\it but also} on a way in which a two-dimensional lattice is deformed into a one-dimensional chain.
As an illustration, a generalized highest weight state  for the staggered ${\rm SU}(3)$ spin-1  ferromagnetic biquadratic model on a $8\times 8$ square lattice, with the periods $p_1=p_2=4$ has been chosen, as shown in  Fig.~\ref{pepsmpsn4}. A PEPS representation is  reshaped into a MPS representation, with the period $p_c$ being 4.   Here the local basis states $|1\rangle$ and $|0\rangle$, which are the eigenvectors of $S^z_p$ at the $p$-th lattice site with the eigenvalues being 1 and 0, are indicated by circles and diamonds, respectively.
We thus are led to a MPS representation for this generalized highest weight state $|{\rm ghws}\rangle_4=|\otimes_{l=1}^{16}\{1110\}_l\rangle$, with the bond dimension being $1$, where the subscript 4 indicates the period. We remark that the period may be identified as the size of the unit cells in the resultant MPS representation.

As a consequence, highly degenerate ground states, for all three models under investigation, admit exact MPS representations, as follows from a prescription for quantum many-body systems undergoing SSB with type-B GMs in one spatial dimension~\cite{smexactmps}. More precisely, the prescription consists of the three steps as follows. The first step is to identify an exact  MPS representation for $|{\rm hws}\rangle$ or $|{\rm ghws}\rangle$. The second step is to figure out an exact MPO representation for  $F_\alpha^{\;M_\alpha}$ ($\alpha=1, 2, \ldots, r$). The third step is to contract the MPO representation for $F_\alpha^{\;M_\alpha}$ with the MPS representation for  $|{\rm hws}\rangle$ or $|{\rm ghws}\rangle$, thus yielding an exact MPS representation for a degenerate ground state  $F_\alpha^{\;M_\alpha}|{\rm hws}\rangle$ or $F_\alpha^{\;M_\alpha}|{\rm ghws}\rangle$.
This offers an efficient way to evaluate the norms for degenerate ground states, which is necessary to perform a finite system-size scaling analysis for the entanglement entropy. Physically, the presence of exact MPS representations reflects the fact that the contribution from the area law to the entanglement entropy is absent, given both the closeness of degrees of freedom to the boundary between a subsystem and its environment and the boundary itself are  not well-defined for highly degenerate ground states arising from SSB with type-B GMs. In other words, the absence of  a contribution from the area law to the entanglement entropy manifests itself in the presence of an exact MPS representation for highly degenerate ground states arising from SSB with type-B GMs.

In principle, the above argument is valid for quantum many-body systems undergoing SSB with type-B GMs in any spatial dimension.

\subsection{The entanglement contour for scale-invariant states arising from SSB with type-B GMs}

Traditionally, SSB with type-B GMs is characterized in terms of an order parameter operator commuting with a Hamiltonian. However, it is this feature that lies at the heart of the debate between Anderson and Peierls regarding whether or not the ${\rm SU}(2)$ ferromagnetic states should be regarded as a result of SSB~\cite{smanderson,smpeierls}. To a certain extent, the introduction of type-A and type-B GMs~\cite{smwatanabe}, which results in the counting rule of GMs, partially resolves this debate, but this does not offer any further characterization of SSB, as far as the differences between type-A and type-B GMs are concerned. In our opinion, the most notable feature of SSB with type-B GMs is that the model Hamiltonians are frustration-free~\cite{smtasaki,smkatsura}. They are thus exactly solvable for highly degenerate ground states. In particular, a permutation symmetry group emerges with respect to the unit cells present in highly degenerate ground states. We stress that the unit cells themselves are {\it emergent}, since they are different from the lattice unit cells~\cite{smexactmps}.

Given the emergence of the unit cells in highly degenerate ground states for a quantum many-body system undergoing SSB with type-B GMs, it is convenient to introduce $j$ and $k$ to label the unit cells. Assume that the lattice is partitioned into a block $B$ and its environment $E$. Then the entanglement contour~\cite{smguifre}, which assigns a real positive number $s_B(j)$ to each of unit cells in a block $B$ such that the sum of $s_B(j)$ over
all the unit cells contained in the block $B$ is equal to the entanglement entropy $S(B)$. Physically, it quantifies how much the degrees of freedom at the unit cell $j$ contribute to the entanglement between the block $B$ and its environment $E$. We remark that, for a finite-size quantum many-body system,  the entanglement entropy $S(B)$ is non-decreasing with the block volume, as long as the block volume is not greater than half the system volume.

Mathematically, the entanglement contour  $s_B(j)$ is subject to the constraints: (i) $s_B(j)$ must be non-negative for any $j$; (ii)  $s_B(j)$ satisfies the normalization  $\sum_{j \in B} s_B(j) = S(B)$; (iii) $s_B(j)$ is symmetric, i.e.,
$s_B(j)=s_B(k)$,  if the reduced density matrix $\rho_B$ is invariant under a symmetry operation that exchanges the unit cell $j$ with the unit cell $k$; (iv)  $s_B(X)$ is invariant under a local unitary transformation, acting on a subset $X$ of the unit cells
contained in the block $B$; and (v) $s_B(X)$ is bounded from above by the entanglement entropy $S(\Omega_B)$: $s_B(X) \leq S(\Omega_B)$, where $X \subseteq B$ is a subset of the unit cells contained in the block $B$, subject to the condition that the Hilbert subspace supported on $X$ is contained in a factor space supported on $\Omega_B$. In particular, the equality is valid if the factor space supported on $\Omega_B$ is identical to a factor space supported on a a set of unit cells in the block $B$. Here, the entanglement contour $s_B(X)$ on a subset $X$ of the unit cells contained in the block $B$, i.e., $ X \subseteq B$, has been defined as the sum of the entanglement contour over the unit cells contained in the subset $X$: $s_B(X) = \sum_{j \in X} s_B(j)$.
It follows that, for any two
subsets $X\subseteq B$ and $Y \subseteq B$ contained in the block $B$, but they are not intersecting with each other: $X \cap Y =\phi$ with $\phi$ being void, then the entanglement contour is additive. That is, we have
$s_B(X\cup Y) = s_B(X) + s_B(Y)$. Meanwhile, if $X \subseteq Y$ are such that all the unit cells of $X$ are contained in $Y$ then the entanglement
contour on $Y$ must not be less than that on $X$, thus establishing the monotonicity of $s_B(X))$, i.e., $s_B(X) \leq s_B (Y)$ if $X \subseteq Y$.

As it turns out, for highly degenerate ground states in quantum many-body systems undergoing SSB with type-B GMs, the entanglement contour $s_B(j)$ is flat, in the sense that every unit cell inside the block $B$ contributes equally to the entanglement entropy $S(B)$, as a consequence of the subgroup of the emergent permutation symmetry with respect to the unit cells (contained in the block $B$). In fact, the symmetric requirement $s_B(j)=s_B(k)$ amounts to establishing an equivalence relation between the two unit cells under an exchange operation $j \leftrightarrow k$, thus inducing an equivalence class among the unit cells. Hence, the presence of the subgroup of the emergent permutation symmetry with respect to the unit cells (contained in the block $B$) implies that there is one and only one equivalence class among the unit cells contained in the block $B$. Mathematically, we have  $s_B(j)= (1/|B|)S(B)$ for any $j \in B$, where $|B|$ denotes the number of unit cells contained in the block $B$. Obviously, this flat entanglement contour $s_B(j)$ satisfy the constraints (i), (ii) and (iii). The constraint (iv) regarding invariance under a local unitary transformation follows from an observation that, generically, such a  local unitary transformation is invariant under the action of the subgroup of the emergent permutation symmetry with respect to the unit cells (contained in the block $B$). As for the constraint (v), note that a degenerate ground state is not factorized, unless it is the highest weight state or a generalized highest weight state with a specific unit cell size, as a result of the action of the lowering operator(s) for the symmetry group $G$. We stress that, in general, the factorization of the highest weight state or a generalized highest weight state is not necessary. Meanwhile, the factorization of the highest weight state or a generalized highest weight state simply means that the corresponding values of both the entanglement contour $s_B(j)$ and the entanglement entropy $S(j)$ vanish for any unit cell $j$.

In other words, quantum many-body systems undergoing SSB with type-B GMs constitute a realization of the simplest spatial fine structure quantified by the entanglement contour.
In this sense, they are {\it trivial} from a perspective of  the entanglement contour, in sharp contrast to those undergoing SSB with type-A GMs in two and higher spatial dimensions and critical quantum many-body systems described by conformal field theory in one spatial dimension.

The ramifications are far-reaching. In fact, the flatness of the entanglement contour explains why the entanglement entropy  $S(B)$ does not depend on what types of boundary conditions are adopted for quantum many-body systems undergoing SSB with type-B GMs, as long as a degenerate ground state under OBCs remains to be a degenerate ground state under PBCs. Indeed, the origin of the logarithmic scaling term in the entanglement entropy for a degenerate ground state arising from SSB with type-B GMs is different from that for a critical ground state described by conformal field theory~\cite{smcft} in one spatial dimension. In fact, for the former, the (additive) logarithmic scaling term is universal, with a contribution from the area law vanishing, thus it {\it only} involves low energy degrees of freedom characterizing the reaction of the gapless type-B GMs to the presence of a length scale when the system itself is partitioned into the block $B$ and the environment $E$.  In contrast, for the latter,  the (multiplicative) logarithmic scaling correction to the area law simultaneously involves both the degrees of freedom near and away from the boundary between the block $B$ and the environment $E$. That is, the origin of the  multiplicative logarithmic correction to the area law may be attributed to the degrees of freedom further away from the boundary. The same picture offers an explanation why the Renyi entropy depends on the Renyi index for  a critical ground state, but not for highly degenerate ground states arising from SSB with type-B GMs.

Our discussion about the spatial fine structure of the entanglement entropy from the entanglement contour reveals a few drastic differences between type-A and type-B GMs, thus deepening the current understanding of  the debate between Anderson and Peierls regarding whether or not the ${\rm SU}(2)$ ferromagnetic states should be regarded as a result of SSB~\cite{smanderson}. In our opinion, it makes sense to speak of SSB with type-B GMs in  the ${\rm SU}(2)$ ferromagnetic model. However, the drastic differences between type-A and type-B GMs, as revealed, clearly indicate that it does not qualify as a paradigmatic example for SSB itself.

\subsection{Eigenvalues of the reduced density matrices for scale-invariant states in the three illustrative models}

As argued in Sec.~A of the SM, the emergent permutation symmetry makes it possible to turn degenerate ground states in two spatial dimensions into those in one spatial dimension for quantum many-body systems undergoing SSB with type-B GMs. One might take advantage of this fact to drastically simplify the evaluation of eigenvalues of the reduced density matrices for scale-invariant states in two spatial dimensions. As it turns out, the expressions take the same form as those for their counterparts in one spatial dimension, which have been derived in Refs.~\cite{FMGM-S,golden-S},  with $L$ being replaced by $L_1L_2$ and $n$ being replaced by $n_1n_2$.

Here we collect the explicit expressions for the eigenvalues of the reduced density matrices  $\rho(L_1,L_2,n_1,n_2) \equiv \rho(L_1L_2,n_1n_2)$ for the
highly degenerate ground states of the three fundamental models on a square lattice.

For the ${\rm SU}(2)$ spin-$s$ ferromagnetic Heisenberg model on the square lattice, the entanglement entropy $S(L_1L_2,n_1n_2,M)$ for a highly degenerate ground state $|L_1L_2,M\rangle$  takes the form
	\begin{align}
	S(L_1L_2,n_1n_2,M)= -\sum_{k=0}^{2sn_1n_2}\Lambda(L_1L_2,n_1n_2,k,M)\log_{2}\Lambda(L_1L_2,n_1n_2,k,M),\label{su2slnf}
\end{align}
where the eigenvalues $\Lambda(L_1L_2,n_1n_2,k,M)$ of the reduced density matrix $\rho(L_1L_2,n_1n_2,M)$ are
\begin{equation}
	\Lambda(L_1L_2,n_1n_2,k,M)=\frac{\mu(L_1L_2,k,M)}{\nu(L_1L_2,k,M)},
\end{equation}
with
	\begin{equation*}
		\mu(L_1L_2,k,M)={\sum}'_{n_{-\!s},\ldots,\; n_{s},\atop l_{-\!s},\ldots,\;l_{s}}\prod_{u,v=-s}^{s-1}\!\varepsilon(s,u)^{n_{u}}
		{C_{n_1n_2-\sum_{m=-s}^{u-1}\!n_m}^{n_u}}\varepsilon(s,v)^{l_{v}} \nonumber \\
		{C_{L_1L_2-n_1n_2-\sum_{m=-s}^{v-1}l_m}^{l_v}},
	\end{equation*}
	and
	\begin{equation*}
		\nu(L_1L_2,k,M)={\sum}'_{N_{-s},\ldots,N_{s}}\prod_{u=-s}^{s-1}\varepsilon(s,u)^{N_{u}}
		{C_{L_1L_2-\sum_{m=-s}^{u-1}N_m}^{N_u}}.
	\end{equation*}
	Here the sum $\sum'_{n_{-s},\ldots,\;n_{s}}$ is taken over all possible values of $n_{-s}$, \ldots , $n_s$, subject to the constraints $\sum_{m=-s}^s n_m=n_1n_2$ and $\sum_{m=-s}^{s}(s-m)n_m=k$, and $\sum'_{l_{-s},\ldots,\;l_{s}}$ is taken over all the possible values of $l_{-s}$, \ldots, $l_s$, subject to the constraints: $\sum_{m=-s}^s l_m=L_1L_2-n_1n_2$.
	The sum $\sum_{m=-s}^{s}(s-m)l_m=M-k$,   $\sum'_{N_{-s},\ldots,\;N_s}$ is taken over all possible values of $N_{-s}$, \ldots, $N_s$, subject to the constraints $\sum_{m=-s}^s N_m=L_1L_2$ and $\sum_{m=-s}^{s}(s-m)N_m=M$.
	The factor $\varepsilon(s,u)$  takes the form
	\begin{equation*}
		\varepsilon(s,u)=\frac{\prod_{m=u+1}^{s}{(s+m)(s-m+1)}}{\prod_{m=u}^{s-1}(s-m)^2}.
	\end{equation*}

For the ${\rm SU}(2s+1)$ ferromagnetic model on the square lattice, the entanglement entropy $S(L_1L_2,n,M_1,\ldots,M_{2s})$ for a highly degenerate ground state $|L_1L_2,M_1,M_2,\ldots, M_{2s}\rangle$   takes the form
	\begin{equation}
		S(L_1L_2,n,M_1,\ldots,M_{2s})= -\sum_{k_1,\ldots,k_N=0}^{n_1n_2}\Lambda(L_1L_2,n_1n_2,k_1,\ldots,k_{2s},M_1,\ldots,M_{2s})\log_{2}\Lambda(L_1L_2,n_1n_2,k_1,\ldots,k_{2s},M_1,\ldots,M_{2s}),\label{sunslnf}
	\end{equation}
	where the eigenvalues $\Lambda(L_1L_2,n_1n_2,k_1,\ldots,k_{2s},M_1,\ldots,M_{2s})$ of the reduced density matrix $\rho(L_1L_2,n_1n_2,M_1,\ldots,M_{2s})$ are
	\begin{align}
		\Lambda(L_1L_2,n_1n_2,k_1,\ldots,k_{2s},M_1,\ldots,M_{2s})=
		\frac{\prod_{\alpha=1}^{2s} {C_{n_1n_2-\sum_{\beta=1}^{\alpha-1}{k_\beta}}^{k_{\alpha}}\prod_{\gamma=1}^{2s} C_{L_1L_2-n_1n_2-\sum_{\beta=1}^{\gamma-1}{(M_\beta-k_\beta)}}^{M_\gamma-k_{\gamma}}}}
		{\prod_{\alpha=1}^N {C_{L_1L_2-\sum_{\beta=1}^{\alpha-1}{M_\beta}}^{M_{\alpha}}}} \;.
	\end{align}

For the staggered ${\rm SU}(3)$ spin-1  ferromagnetic biquadratic model on the square lattice, we consider two highly degenerate ground states $|L_1L_2,M_1,M_2\rangle_{22}$ with the periods $q_1=q_2=2$, and $|L_1L_2,M_2,M_3\rangle_{44}$ with the periods $q_1=q_2=4$.
To keep consistency, $|L_1L_2,M_2,M_3\rangle_{44}$ may be denoted as $|L_1L_2,M_1^*,M_2^*\rangle_{44}$, where the fillings  $f_1^*$ and  $f_2^*$ are defined as $f_1^*=M_1^*/L$ and $f_2^*=M_2^*/L$, with $M_1^*=M_2$ and $M_2^*=M_3$.
The entanglement entropy $S(L_1L_2,n_1n_2,M_1,M_2)$  takes the form
	\begin{equation}
		S(L_1L_2,n_1n_2,M_1,M_2)= -\sum\limits_{k_1=0}^{\min(M_1,n_1n_2/2)}\sum\limits_{k_2=0}^{\min(M_2,n_1n_2-k_1)}\Lambda(L_1L_2,n_1n_2,k_1,k_2,M_1,M_2)\log_{2}\Lambda(L_1L_2,n_1n_2,k_1,k_2,M_1,M_2),\label{stsu3slnf1f2}
	\end{equation}
	where both $n_1$ and $n_2$ are a multiple of two and the eigenvalues $\Lambda(L_1L_2,n_1n_2,k_1,k_2,M_1,M_2)$ of the reduced density matrix $\rho(L_1L_2,n_1n_2,M_1,M_2)$ are
	\begin{equation}
		\Lambda(L_1L_2,n_1n_2,k_1,k_2,M_1,M_2)=\frac{C_{n_1n_2/2}^{k_1}C_{n_1n_2-k_1}^{k_2}C_{(L_1L_2-n_1n_2)/2}^{M_1-k_1}C_{L_1L_2-n_1n_2-M_1+k_1}^{M_2-k_2}} {C_{L_1L_2/2}^{M_1}C_{L_1L_2-M_1}^{M_2}}.
		\label{lamm1m2}
	\end{equation}
	The entanglement entropy $S(L_1L_2,n_1n_2,M_1^*,M_2^*)$  takes the form
	\begin{equation}
		S(L_1L_2,n_1n_2,M_1^*,M_2^*)= -\sum\limits_{k_2=0}^{n_1n_2/4}\sum\limits_{k_1=0}^{3n_1n_2/4-k_2}\Lambda(L_1L_2,n_1n_2,k_1,k_2,M_1^*,M_2^*)\log_{2}\Lambda(L_1L_2,n_1n_2,k_1,k_2,M_1^*,M_2^*),\label{stsu3slnf2f3}
	\end{equation}
where both $n_1$ and $n_2$ are a multiple of four, and the eigenvalues $\Lambda(L_1L_2,n_1n_2,k_1,k_2,M_1^*,M_2^*)$ are
\begin{equation}
	\Lambda(L_1L_2,n_1n_2,k_1,k_2,M_1^*,M_2^*)=\frac{C_{n_1n_2/4}^{k_2}C_{3n_1n_2/4-k_2}^{k_1}C_{(L_1L_2-n_1n_2)/4}^{M_2^*-k_2}C_{3(L_1L_2-n_1n_2)/4-M_2^*+k_2}^{M_1^*-k_1}} {C_{L_1L_2/4}^{M_2^*}C_{3L_1L_2/4-M_2^*}^{M_1^*}}.
	\label{lamm2m3}
\end{equation}

Here we remark that our numerical data have been presented for a particular choice $L_1 = L_2$ and $n_1= n_2$ in the main text, which is sufficient to extract the number of type-B GMs $N_B$.

\end{document}